\documentclass{aa520}
\usepackage{graphicx,txfonts}

\begin{document}

\title{Stability of rotating spherical stellar systems}

\author{Andr\'es Meza\thanks{Present Address: Department of Physics and
Astronomy, University of Victoria, Victoria BC, V8P 1A1, Canada. E-mail:
ameza@uvic.ca.}}

\institute{Departamento de Astronom\'{\i}a y Astrof\'{\i}sica, Pontificia
Universidad Cat\'olica de Chile, Casilla 306-22, Santiago, Chile}

\date{Received / Accepted}

\abstract{ 
The stability of rotating isotropic spherical stellar systems is investigated by
using N-body simulations. Four spherical models with realistic density profiles
are studied: one of them fits the luminosity profile of globular clusters, while
the remaining three models provide good approximations to the surface brightness
of elliptical galaxies. The phase-space distribution function $f(E)$ of each one
of these non-rotating models satisfies the sufficient condition for stability
$df/dE<0$.  Different amounts of rotation are introduced in these models by
changing the sign of the $z$-component of the angular momentum for a given
fraction of the particles. Numerical simulations show that all these rotating
models are stable to both radial and non-radial perturbations, irrespective of
their degree of rotation. These results suggest that rotating isotropic spherical
models with realistic density profiles might generally be stable. Furthermore,
they show that spherical stellar systems can rotate very rapidly without becoming
oblate.
\keywords{celestial mechanics, stellar dynamics -- galaxies: kinematics and
dynamics -- instabilities -- methods: n-body simulations}}

\maketitle

\section{Introduction}

Dynamical instabilities in spherically symmetric stellar systems have been
investigated for more than four decades. In a seminal work, Antonov
(\cite{antonov60,antonov62}) used a variational principle to demonstrate that
non-rotating spherical models with a phase-space distribution function $f$
depending only on the energy $E$ are stable to non-radial perturbations if
$df/dE<0$. Subsequent works showed that this condition is also a sufficient
condition for stability to radial perturbations (Dor\'emus et al.
\cite{doremus71}; Sygnet et al. \cite{sygnet84}; Kandrup \& Sygnet
\cite{kandrup85}). In general, non-rotating spherical stellar systems are
described by distribution functions $f$ that depend on both the energy $E$ and the
magnitude of the angular momentum $L$. In such systems only the stability to
radial modes can be tested by using the sufficient condition $\partial f/\partial
E<0$ (Dor\'emus \& Feix \cite{doremus73}; Dejonghe \& Merritt \cite{dejonghe88}).
For this reason, numerical simulations have been an indispensable tool to
investigate the stability of anisotropic spherical models. 

Several classes of instabilities have been discovered in non-rotating spherical
models with anisotropic velocity distributions (e.g., H\'enon \cite{henon73};
Merritt \& Aguilar \cite{merritt85a}; Barnes et al. \cite{barnes86}; see Merritt
\cite{merritt99} for a recent review). For example, models dominated by stars on
radial or eccentric orbits can be unstable to forming a triaxial bar (Polyachenko
\cite{polyachenko81}; Merritt \& Aguilar \cite{merritt85a}; Meza \& Zamorano
\cite{meza97}), while models composed mainly of stars on circular orbits can
exhibit quadrupole-mode oscillations (Barnes et al. \cite{barnes86}).  

In contrast to non-rotating spherical models, comparatively little work has been
done to investigate the stability of rotating spherical stellar systems. Miller \&
Smith (\cite{miller80}) employed several methods to introduce rotation in a
spherical $n=3$ polytrope with isotropic velocity distribution. In all cases, they
found that the addition of rotation does not affect the stability of their models;
indeed, they found that even rapidly rotating systems remain spherically
symmetric. More recently, Alimi et al. (\cite{alimi99}) showed that rotating
isotropic spherical $n=4$ polytropes, which were made to rotate by changing the
sign of the $z$-component of the angular momentum for a fraction of the particles,
were dynamically stable.

On the other hand, Allen et al. (\cite{allen92}) reported the existence of a
``tumbling bar instability'' in a set of rotating spherical models with different
degrees of velocity anisotropy, ranging from completely circular to entirely
radial (see also Papaloizou et al. \cite{papaloizou91}; Palmer
\cite{palmer94a,palmer94b}). In particular, they suggested that the introduction
of a small amount of rotation in the isotropic spherical $n=2$ polytrope induces a
bar instability in this otherwise stable system. However, Sellwood \& Valluri
(\cite{sellwood97}) using the same files of initial conditions but a different
N-body code, showed that these models are actually stable; the instability appears
do not exist. They suggested that the evolution observed by Allen et al.
(\cite{allen92}) was, probably, caused by an improper treatment of variable time
steps in their N-body code.

Most of the previous results have been obtained for rotating spherical models with
somewhat unrealistic properties. Indeed, most of these works have employed
spherical polytropes with finite radius (e.g., Alimi et al. \cite{alimi99}) or
models with unrealistic velocity distributions (e.g., Allen et al. \cite{allen92}).
Therefore, they cannot be considered as general. To investigate the influence of
rotation on the stability of spherical stellar systems, it is necessary to
consider models with more realistic density profiles and velocity distributions.
In this paper, the results of a series of N-body simulations for four of such
models are presented. These isotropic spherical models are: the Plummer
(\cite{plummer11}) model, the Hernquist (\cite{hernquist90}) model, the Jaffe
(\cite{jaffe83}) model, and the $\gamma=0$ model (Dehnen \cite{dehnen93}; Tremaine
et al. \cite{tremaine94}). Different degrees of rotation were introduced in these
models by using the so-called Lynden-Bell's (\cite{lynden-bell60,lynden-bell62})
demon to reverse the sense of rotation along the $z$-axis of a given fraction of
the particles. Numerical simulations show that all these rotating models are
stable, regardless of their degree of rotation. 

This paper is organized as follows. The method employed to introduce net rotation
in these models and the N-body code used to follow their dynamical evolution are
described in Sect.~2. The main results of the numerical simulations are summarized
in Sect.~3. Finally, a brief discussion of the results is given in Sect.~4.

\section{Numerical simulations}

\subsection{Rotating models \label{models}}

The spherical models studied in this paper are: the Plummer (\cite{plummer11})
model,
\begin{equation}
\rho = \frac{3}{4\pi} \frac{Ma^2}{(a^2+r^2)^{5/2}}, \end{equation}
the Jaffe (\cite{jaffe83}) model,
\begin{equation}
\rho = \frac{1}{4\pi} \frac{Ma}{r^2(a+r)^2},
\end{equation}
the Hernquist (\cite{hernquist90}) model,
\begin{equation}
\rho = \frac{1}{2\pi} \frac{Ma}{r(a+r)^3},
\end{equation} 
and the $\gamma=0$ model (Dehnen \cite{dehnen93}; Tremaine et al.
\cite{tremaine94}),
\begin{equation}
\rho = \frac{3}{4\pi} \frac{aM}{(a+r)^4}.
\end{equation}
In all these cases, $M$ is the total mass and $a$ is the scale radius. Hereafter,
all quantities are expressed in units such that $M=a=G=1$, where $G$ is the
gravitational constant. Some parameters for these models are summarized in
Table~\ref{table1}, where $r_h$ is the half-mass radius and $t_h$ is the dynamical
time evaluated at $r_h$.

\begin{table}
\caption{Parameters for the models \label{table1}}
\begin{center}
\begin{tabular}{lcrl}
\hline
Model      & $r_{h}$ & $t_{h}$ & $\Delta t$ \\
\hline
Plummer    & 1.31    &  3.32   & 0.008      \\
Hernquist  & 2.41    &  8.33   & 0.02       \\
Jaffe      & 1.00    &  2.22   & 0.006      \\
$\gamma=0$ & 3.85    & 16.78   & 0.04       \\
\hline
\end{tabular}
\end{center}
\end{table}

The Plummer model fits the light distribution of globular clusters (see e.g.,
Spitzer \cite{spitzer87}), while the remaining three density profiles provide good
approximations to the surface brightness of elliptical galaxies (see Dehnen
\cite{dehnen93}). Several intrinsic properties and projected quantities of these
models can be obtained analytically, e.g., the velocity dispersions, the surface
brightness profile and the mass distribution. In particular, the phase-space
distribution function $f(E)$ can be obtained analytically by using the Eddington's
inversion formula (see e.g., Binney \& Tremaine \cite{binney87}). For all these
non-rotating models, the distribution function satisfies the sufficient condition
for stability $df/dE<0$. Therefore, the models are stable to both radial and
non-radial perturbations (Antonov \cite{antonov62}; Doremus et al.
\cite{doremus71}; Sygnet et al. \cite{sygnet84}; Kandrup \& Sygnet
\cite{kandrup85}).

Initial conditions for the simulations were derived from the density profile and
the distribution function by using the following scheme. The radial coordinate of
each particle is assigned by inverting the equation $M(r)=x$, where $M(r)$ is the
total mass inside the radius $r$ and $x$ is a uniform random variable in the range
$[0,1)$. Then, the coordinates of the position vector ${\bf x}$ are chosen at
random from a sphere with radius $r$. The distribution function $f(E)$ and the
gravitational potential $\Phi(r)$ evaluated at the particle position define the
distribution of velocities, which is sampled by an acceptance-rejection technique
to assign the modulus of the velocity $v$ of each particle (see e.g., Press et al.
\cite{press92}). Finally, the coordinates of the velocity are chosen at random
from a sphere with radius $v$.

The distribution function $f(E)$ does not depend on the sign of the $z$-component
of the angular momentum $L_z$ (or equivalently on the sign of $v_\phi$), therefore
the models have no net streaming. Different amounts of rotation can be introduced
in these models by reversing the sense of rotation about some axis, here taken to
be the $z$-axis, of a given fraction of the particles (Lynden-Bell
\cite{lynden-bell60,lynden-bell62}). In a non-rotating model, there are equal
number of particles with $v_\phi$ going in opposite directions, while for a
maximally streaming model the sign of $v_\phi$ is the same for all particles. All
intermediate cases have varying fractions of particles with $v_\phi$ going in
opposite directions. This scheme preserves the position and the norm of the
velocity of each particle, then the systems are put in rotation without modifying
their total kinetic and potential energy. Therefore, the resulting rotating models
are also in dynamical equilibrium (see Lynden-Bell \cite{lynden-bell60}).

This flipping rule introduces a discontinuity in the distribution function across
$L_z=0$, which, in principle, can enhance the strength of the bar mode (Kalnajs
\cite{kalnajs77}). However, as it is shown below, no signs of bar instabilities are
observed in these rotating models. Therefore, no special procedure was
necessary to taper this flipping rule when $|L_z|$ is small.

The degree of rotation in these models can be measured by the parameter (see
e.g., Sellwood \& Valluri \cite{sellwood97})
\begin{equation}
\eta = \frac{\sum_{i=1}^{N} L_{z_{i}}}{\sum_{i=1}^{N} |L_{z_{i}}|},
\end{equation}
which varies from $\eta=0$ for a non-rotating model to $\eta=1$ for a model with
all particles orbiting in the same sense around the $z$ axis. When $\eta=0.5$ the
system has half the maximum possible total angular momentum; in this case, 75\% of
the particles are orbiting in the direct sense and the remaining 25\% are
retrograde.

Alternatively, the amount of rotation can also be given in terms of the following
parameter (see Navarro \& White \cite{navarro93})
\begin{equation}
\mu = \frac{K_\mathrm{rot}}{K},
\end{equation}
where $K$ is the total kinetic energy and $K_{\rm rot}$ is the rotation kinetic
energy defined by
\begin{equation}
K_\mathrm{rot} = \frac{1}{2} \sum_{i=1}^{N} \frac{m_i({\bf L}_i\cdot\hat{\bf
L}_{\rm tot})^2}{r_{i}^{2}-({\bf r}_i\cdot\hat{\bf L}_{\rm tot})^2}
\label{eq:mu}
\end{equation}
In order to exclude counterrotating particles, the sum in equation (\ref{eq:mu})
is carried out only over particles that satisfy the condition $({\bf
L}_i\cdot\hat{\bf L}_\mathrm{tot})>0$, where $\hat{\bf L}_\mathrm{tot}$ is a unit
vector in the direction of the total angular momentum of the system, in this case,
along the $z$-axis. Therefore, this parameter measures the kinetic energy in the
rotational motion around the $z$-axis. It varies from $\mu=1/6$ for a non-rotating
model to $\mu=1/3$, the maximum value allowed for the virial theorem, for a model
with all particles rotating in the same sense around the $z$-axis. Table
\ref{table2} summarizes the values of $\eta$ and the associated values of $\mu$
employed in the simulations.

\begin{table}
\caption{Parameters for the rotating models. \label{table2}}
\begin{center}
\begin{tabular}{cc}
\hline
$\eta$ & $\mu$ \\
\hline
0   & 0.17 \\
0.2 & 0.20 \\
0.4 & 0.23 \\
0.6 & 0.27 \\
0.8 & 0.30 \\
1.0 & 0.33 \\
\hline
\end{tabular}
\end{center}
\end{table}

\begin{figure}[ht]
\centering
\includegraphics[width=\linewidth,clip]{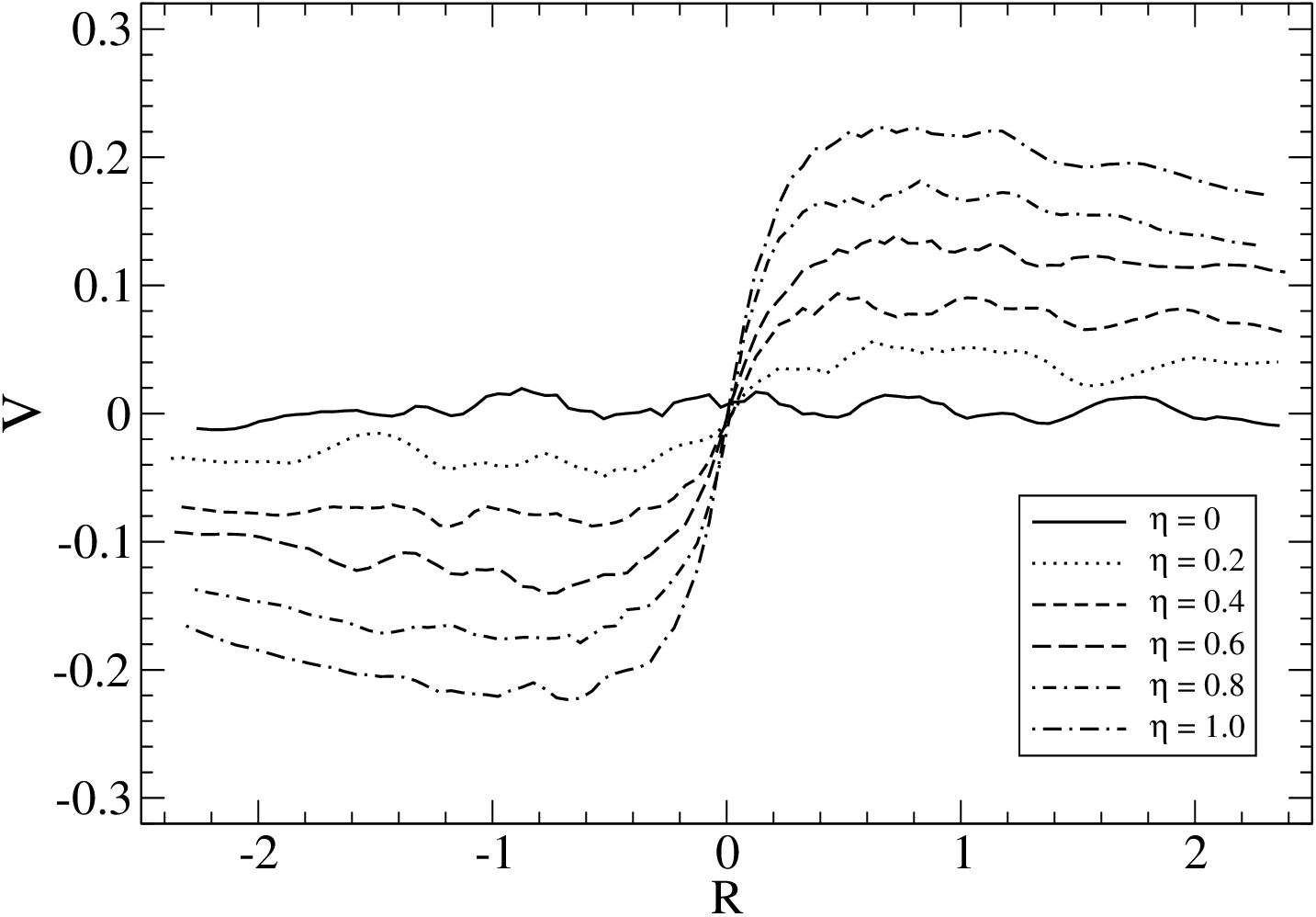}
\caption{Mean streaming velocity for Plummer models with different degrees of
rotation $\eta$ as a function of the projected position along the $x$-axis. The
systems are viewed such that the rotation axis is perpendicular to the
line-of-sight. \label{figure1}}
\end{figure}

Figure \ref{figure1} shows the mean streaming velocity for Plummer models with
different degrees of rotation. To measure this quantity, the systems were first
projected along the $y$-axis and, then, a slit was placed along the projected
$x$-axis. The slit width was set at $r_h$, the half-mass radius, and the bin size
along the projected axis was varied such that each bin contains the same number of
particles ($n_\mathrm{bin}=1000$). In all the rotating models, the velocity curve
reaches the maximun at $R\la r_h$ and then decreases slowly (models with $\eta\ga
0.6$) or remains practically flat (models with $0<\eta\la 0.6$). As expected, the
mean streaming velocity of the non-rotating model ($\eta=0$) is nearly zero; the
observed fluctuations reflect the discreteness of the models. The velocity curves
for the other models are very similar.

\subsection{N-body code}

The stability of these models was investigated by using an N-body code based on
the self-consistent field method described by Hernquist \& Ostriker
(\cite{hernquist92}). This scheme consists of solving the Poisson equation by
expanding the density and the gravitational potential using a biortoghonal set of
basis functions. For spherically symmetric systems, it is natural to employ
spherical harmonics to expand the angular dependence. Then, the density and
potential expansions become
\begin{eqnarray}
\rho({\bf r}) & = & \sum_{nlm} A_{nlm} \rho_{nl}(r)\, Y_{lm}(\theta,\varphi),
\label{eq:density} \\
&&\nonumber \\
\Phi({\bf r}) & = & \sum_{nlm} A_{nlm} \Phi_{nl}(r)\, Y_{lm}(\theta,\varphi).
\label{eq:potential}
\end{eqnarray}
In practice, these expansions are truncated at some values $n_\mathrm{max}$ and
$l_\mathrm{max}$ for the radial and angular functions, respectively.

The choice of the radial basis functions $\{\rho_{nl},\Phi_{nl}\}$ is not unique; in
fact, several sets have been proposed (e.g., Clutton-Brock \cite{clutton-brock73};
Allen et al. \cite{allen90}; Zhao \cite{zhao96}). However, the efficiency of this
method relies upon the ability to represent the density profile of the initial system
and its subsequent evolution with the first few terms of the basis set. Therefore, it
is generally desirable to select a set of basis functions such that their lowest
order terms provide a good approximation to the density profile of the system
under study.

In these simulations, the Clutton-Brock (\cite{clutton-brock73}) basis set is used
for the Plummer model, while the Hernquist-Ostriker (\cite{hernquist92}) basis set
is employed for the other three models. The zeroth-order terms of these basis sets
are, respectively, the Plummer model and the Hernquist model. The $\gamma=0$ model
can be recovered by a linear combination of only two terms of the
Hernquist-Ostriker basis functions (see Hernquist \& Ostriker \cite{hernquist92}).
Therefore, these basis sets provide exact representations for three of the models
here studied. On the other hand, the Jaffe model can only be approximated by using
a finite number of terms of the Hernquist-Ostriker basis set; in this case, good
accuracy can be obtained with $\sim 10$ terms (see e.g., Meza \& Zamorano
\cite{meza97}).

\begin{figure*}[ht]
\centering
\includegraphics[width=0.8\linewidth,clip]{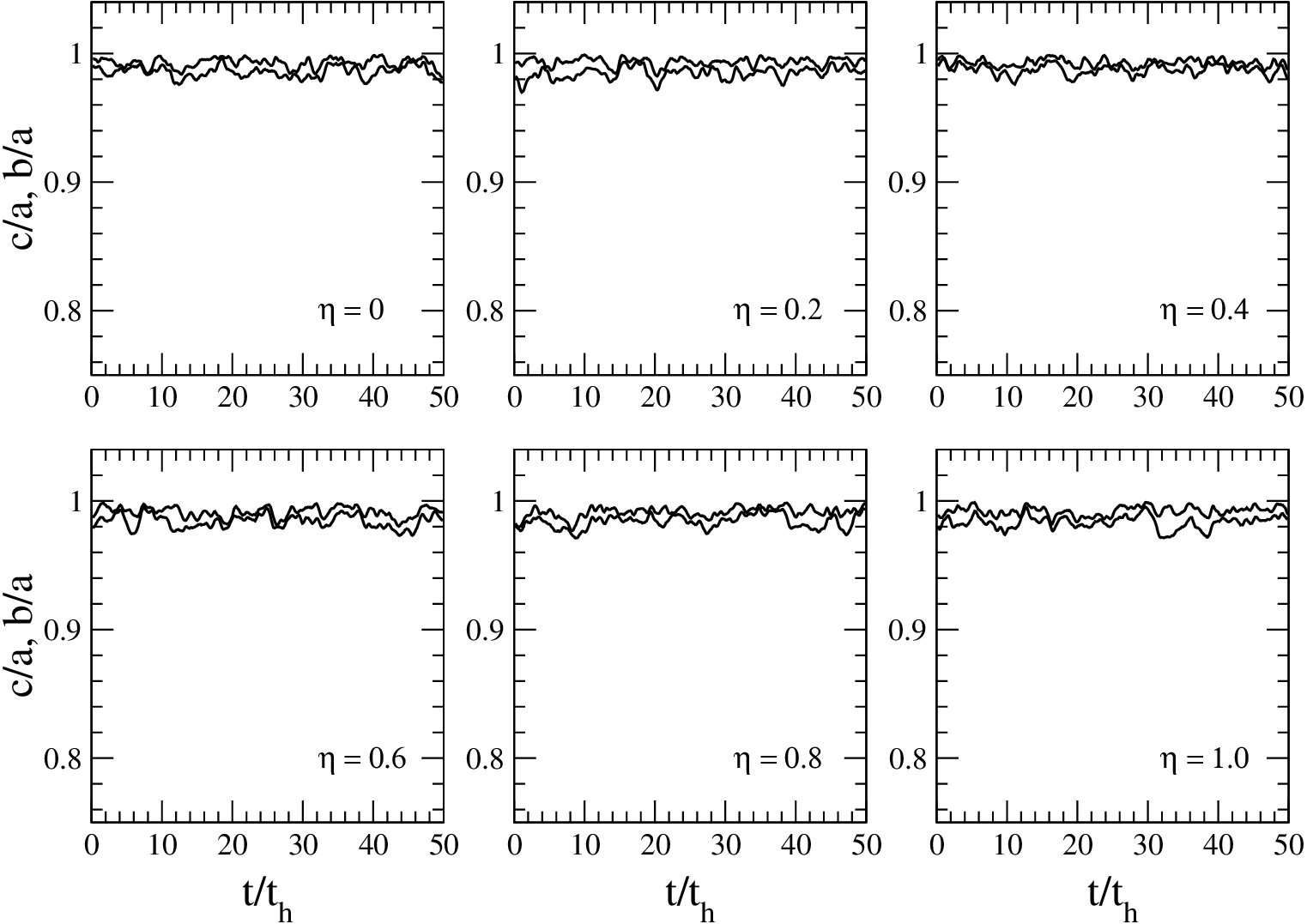}
\caption{Evolution of the axis ratios $c/a$ and $b/a$, with $a\ge b\ge c$, for the
$\gamma=0$ models with different degrees of rotation. The axis ratios were
measured at radius $r_{m}=8$, which encloses $\sim 70\%$ of the total mass. Time
has been normalized to the half-mass dynamical time, $t_{h}=16.8$.
\label{figure2}}
\end{figure*}

All simulations employed $N=10^5$ equal-mass particles and potential expansions up
to $n_\mathrm{max}=10$ for the radial functions and $l_\mathrm{max}=2$ for the
angular functions. Nonzero $m$ terms were also included to allow the development
of any non-axisymmetric instabilities. Time integration was performed using a
second order scheme with a fixed time step $\Delta t$, given by
\begin{eqnarray}
{\bf x}_{i+1} & = & {\bf x}_{i} + \Delta t \, {\bf v}_{i} +
\frac{1}{2} \,  \Delta t^{2} \, {\bf a}_{i}, \\
&&\nonumber \\
{\bf v}_{i+1} & = & {\bf v}_{i} + \frac{1}{2} \, \Delta t \,
({\bf a}_{i} + {\bf a}_{i+1}),
\end{eqnarray}
where the subscript identifies the iteration (see e.g., Hut et al. \cite{hut95}).
The total elapsed time for all these simulations was $T_\mathrm{end} = 50 t_{h}$,
where $t_{h}$ is the dynamical time evaluated at the half-mass radius of the
respective model. The time step used for each model is shown in
Table~\ref{table1}. With these parameters, the total energy was conserved to
better than 0.01\% for the Plummer, Hernquist and $\gamma=0$ models, while for the
Jaffe models the conservation of energy was only about 4\%.

With the adopted value $l_\mathrm{max}=2$ for the potential expansions, these
simulations are designed for searching radial ($m=0$), lopsided ($m=1$), and bar
($m=2$) instabilities. A test particularly sensitive to the bar instability is the
evolution of the axis ratios of the particle distribution. For each simulation,
the axial ratios inside a given radius were obtained by using an iterative
algorithm similar to that used by Dubinski \& Carlberg (\cite{dubinski91}). In
this scheme, initial values for the inertia tensor
\begin{equation}
I_{ij} = \sum \frac{x_i x_j}{|{\bf x}|^2}.
\end{equation}
are computed for all particles inside a sphere of radius $r_m$. The eigenvalues
and eigenvectors of $I_{ij}$ provide an approximation to the axis ratios and the
orientation of the fitting ellipsoid. Then, the modified inertia tensor is
evaluated only for particles inside that ellipsoid, which gives an improved
approximation to their axis ratios and orientation. This process is repeated until
the axis ratios converge to a value within a pre-established tolerance.

\section{Results}

A total of 24 simulations were performed. The main result of these simulations is
that all the rotating models are dynamically stable. This is illustrated by
displaying the evolution of some of them. The results for other models are
similar.

The evolution of the axis ratios $b/a$ and $c/a$, where $a\ge b\ge c$, for the
$\gamma=0$ models with different degrees of rotation is shown in Figure
\ref{figure2}. These axis ratios were measured at radius $r_m=8$, which
encloses $\sim 70\%$ of the total mass. For all values of $\eta$, the axis
ratios remains essentially equal to their initial values. The fluctuations
reflect the discreteness of the models and are consistent with the number of
particles employed in the simulations. Similar behavior is observed for the
axis ratios measured at other radii.

\begin{figure*}[ht]
\centering
\includegraphics[width=0.9\linewidth,clip]{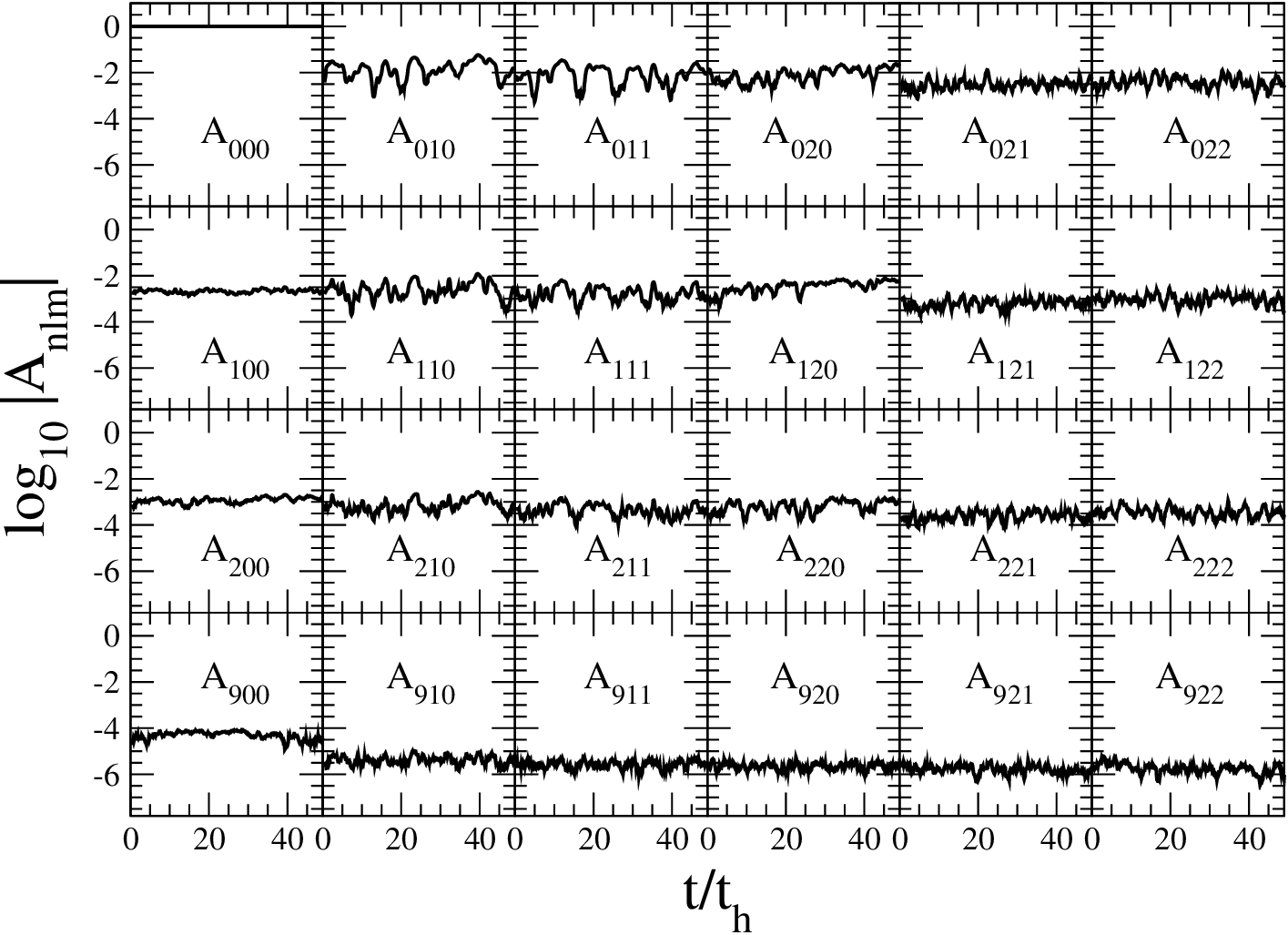}
\caption{Evolution of the amplitude of several expansion coefficients $A_{nlm}$
for the Hernquist model with $\eta=1$. The coefficients $A_{nlm}$ were computed
using the Hernquist-Ostriker basis set, where $n$ refers to the radial functions,
and $l$ and $m$ to the angular functions (spherical harmonics). The logarithm of
the absolute value of the coefficients is shown. Time has been normalized to the
half-mass dynamical time, $t_{h}=8.33$. \label{figure3}}
\end{figure*}

A more sensitive test for the existence of possible instabilities is provided by
the analysis of the individual expansion coefficients $A_{nlm}$, which is
straightforward in the self-consistent field method because the coefficients are
evaluated at each time step in order to compute the forces. The evolution of the
amplitude of several expansion coefficients $A_{nlm}$ for the rotating Hernquist
model with $\eta=1$ is shown in Figure \ref{figure3}. These coefficients were
calculated using the Hernquist \& Ostriker (\cite{hernquist92}) basis functions.
As expected, the zeroth-order coefficient $A_{000}\simeq 1$, while the higher
order coefficients are closer to zero and fluctuate around their initial values
with amplitude consistent with the noise induced by the number of particles used
in these simulations. In particular, there are no signs of radial ($m=0$), bar
($m=2$), or lopsided ($m=1$) instabilities. Other coefficients show the same
qualitative behavior.

\section{Conclusions}

The stability of four rotating isotropic spherical models was investigated by using
N-body simulations. The density profiles of these models provide good
approximations to the mass distribution of globular clusters and elliptical
galaxies. Different degrees of rotation were introduced in these models by
reversing the sense of rotation along the $z$-axis of a given fraction of the
particles (Lynden-Bell \cite{lynden-bell62}). Simulations show that all these
rotating models are dynamically stable, irrespective of their degree of rotation.
No signs of radial, lopsided, or bar instabilities were observed. If some of them
exist their growth rate is larger than $50t_h$, the time elapsed for these
simulations.

These simulations show that spherical stellar systems can rotate very rapidly
without becoming oblate. This result contrasts with the suggestion of Alimi et al.
(\cite{alimi99}) that there do not exist spherical stellar systems in fast
rotation (i.e., with $\mu\ga 0.1$). However, their conclusions were based on a
series of simulations for spherical $n=4$ polytropes with isotropic velocity
distributions, which were made to rotate by using a procedure that modifies their
initial velocity anisotropy. Therefore, their conclusions are not strictly
applicable to the isotropic models studied in this paper. But, they could still be
valid for anisotropic spherical models. In fact, there are several other ways to
construct rotating spherical models. For example, one could construct a system
containing only circular orbits and then reverse a half of them (see Lynden-Bell
\cite{lynden-bell60}). Clearly, such a model would exhibit stronger rotation that
the models here analyzed and surely might be unstable.

\begin{acknowledgements}
I am grateful to the referee for providing comments which improved the
presentation of this paper. I thank Andreas Reisenegger and Nelson Zamorano for
helpful suggestions and comments on the manuscript. I also thank the hospitality
of the Department of Physics and Astronomy at the University of Victoria where the
final version of this paper was written. This work was partially supported by
FONDECYT grant 3990031. 
\end{acknowledgements}

\end{document}